\documentclass[11pt]{article}
\usepackage{geometry}
\usepackage{authblk}
\usepackage{setspace}
\usepackage{graphicx}
\usepackage{dcolumn}
\usepackage{bm}
\usepackage{epsfig,amsmath}
\usepackage{amssymb}
\usepackage{graphics}
\usepackage{latexsym}
\usepackage{placeins}
\usepackage{tikz}
\usepackage{mathrsfs}
\usepackage{subfigure,esint}
\usepackage{float}
\usepackage{color}
\usepackage[pagebackref=false, colorlinks=true]{hyperref}
\definecolor{redish}{rgb}{0.7,0.2,0.0}  
\definecolor{bluish}{rgb}{0.2,0.5,0.8}

\hypersetup{linkcolor=redish,          
                  citecolor=blue,        
                  filecolor=magenta,      
                  urlcolor=blue}          

\begin{document}
\date{}
\author{Suvankar Paul \thanks{\href{mailto:svnkr@iitk.ac.in}{svnkr@iitk.ac.in}}}
\affil{\it Department of Physics, Indian Institute of Technology, Kanpur 208016, India}

\title{\Large\bfseries Strong gravitational lensing by a strongly naked null singularity}

\maketitle

\begin{abstract}
We study strong gravitational lensing in a static, spherically symmetric, naked singularity spacetime, without a 
photon sphere. The nature of the singularity is found to be lightlike. We discuss the characteristic lensing features of this naked 
singularity in the strong deflection limit. In spite of the absence of a photon sphere in this spacetime, the bending angle 
of light diverges, as it approaches the singularity. However, unlike black holes, it is found that the nature of this 
divergence is nonlogarithmic, and we derive an analytic formula for the same.
Moreover, the relativistic rings 
produced due to strong lensing by the singularity are found to be well separated from each other, making them 
easy to resolve and possibly detect. These features are expected to be important in the study of strong lensing by 
ultracompact objects, especially ones without event horizons.
\end{abstract}
\maketitle


\section{Introduction}
General relativity (GR) is one of the most successful theories of gravity to date. It has so far passed many experimental tests,
of which the deflection of light in a gravitational field is one of the most significant. In fact, this was the very first experimental test 
that was used to validate GR more than a century ago, in 
1919 \cite{Dyson-1920}. Most of these tests deal with gravity in the weak field 
limit --- for example, gravity around the Sun or planets in our Solar System. With the advent of modern technology, it
now looks possible to probe deeper --- i.e., into the strong gravity regime. 
Indeed, the Event Horizon Telescope, and the recent imaging of the center of the 
M87 galaxy (\cite{Akiyama-2019a}-\cite{Akiyama-2019f}), 
along with the recent detection of gravitational waves by the LIGO-VIRGO Collaboration 
(see, for example, Refs. \cite{Abbott-2016a,Abbott-2016b}), 
has started a new era in observational astronomy to probe GR in regions of strong gravity.

Ubiquitous in such studies of strong gravity are black holes --- singularities covered by an event horizon. 
It is commonly believed that the centers of galaxies are inhabited by supermassive black holes. 
This has led to a flurry of activities in the recent past which concern gravitational lensing in black hole 
backgrounds. Close to a black hole horizon, gravity might become strong, and it is imperative that here 
one goes beyond the weak field limit to understand lensing from these objects. 
In recent years such studies have been extended to objects that do not have an event horizon. One such
spacetime that has received a lot of attention of late is that of a naked singularity. 
Although the cosmic censorship conjecture prohibits the formation of naked singularities \cite{Penrose-2002}, 
a plethora of recent studies in the literature have shown that it is possible to form such singularities as the 
end product of the gravitational collapse of massive objects under suitable initial conditions
(\cite{Joshi-1993}-\cite{Harada-2998}). Therefore, the study 
of naked singularities, or more generally, horizonless compact objects, is of interest. In this regard, a natural question 
that arises is how to distinguish such objects from black holes. Once again, gravitational lensing in the strong 
deflection limit acts as a significant tool to observationally distinguish between black holes and naked singularities, 
or horizonless objects. The pioneering work on gravitational lensing from naked singularities
by Virbhadra and Ellis \cite{Virbhadra-2002} appeared almost two decades ago. 
Ever since, different aspects of gravitational lensing by various horizonless objects, such 
as naked singularities (\cite{Virbhadra-2008}-\cite{Dey-2020}), wormholes
(\cite{Cramer-1995}-\cite{Wang-2020}), 
Bosonic stars \cite{Cunha-2017}, gravastars \cite{Kubo-2016}, superspinars \cite{Bambi-2019} (objects described by the Kerr metric having a rotation parameter, $a>M$ \cite{Gimon-2009}), etc., have been studied and analyzed in detail. 
Recently, there have also been many other studies on the observational aspects of ultracompact 
objects (\cite{Shaikh-2019c}-\cite{Cunha-2018}) in the context of strong lensing. 
It has been observed from these studies that, while such objects may sometimes produce strong lensing features 
mimicking those of black holes, they might also differ dramatically from black holes, thus opening a possibility for their 
observational detection.

In this paper, we extend such analysis and consider a naked singularity recently proposed by Joshi {\it et} al. in Ref. \cite{Joshi-2020}, 
and we study its lensing features in the strong deflection limit. A curvature singularity (with or without an event horizon) can, 
in general, be classified into three categories --- namely spacelike, timelike, and lightlike or null. For example, the Schwarzschild 
singularity is spacelike, the Reissner-Nordstrom (RN) singularity (for both the black hole and naked singularity cases) 
is timelike, and for references on null singularity, one may refer to Refs. \cite{Marlof-2012,Scheel-2014}.
Null singularities are generally formed at late times inside the RN or Kerr black holes. The inner Cauchy horizons 
(which are inside the outer event horizons) in these black holes
usually break up into two kinds of null singularities as the black holes become old \cite{Marlof-2012,Scheel-2014}.
The interesting feature of the singularity that we consider in this paper is that it is found to be both null and naked. 
Moreover, as is well known, we can further classify a naked singularity 
based on whether it is cloaked by a photon sphere or not \cite{Virbhadra-2002}. If a naked singularity is within 
a photon sphere, it is called a weakly naked singularity (WNS); otherwise, it is termed as a strongly naked singularity (SNS). 
They have their own characteristic strong lensing features, different from each other.
A WNS generally produces strong lensing features similar to those of black holes --- i.e., the bending angle diverges at 
some critical limit in both cases. A WNS can thus in principle act as a black hole mimicker. By contrast, an SNS
that does not have any photon sphere is expected to show crucial differences from a WNS or a black hole. 
The naked singularity spacetime that we consider in this paper also does not contain a photon sphere \cite{Joshi-2020}, so 
it falls in the SNS category. Remarkably, what we will show below is that the bending angle still diverges at a specific 
critical limit, just like a WNS or a black hole. Further, it is well known that for strong lensing due to photon spheres (i.e., in black holes or WNSs), 
the characteristic divergence of the bending angle is logarithmic. Here, on the other hand, we show that this divergence is 
nonlogarithmic, and we derive an analytic formula for the same. 
These are novel results on strong lensing from a naked null singularity, which will be established in
the rest of the paper.

This paper is organized as follows: in Sec. \ref{sec:properties}, we recall some features 
of the spacetime that we study. Then we briefly discuss the properties of null geodesics in this spacetime 
in Sec. \ref{sec:photon-motion}. Section \ref{sec:strong-bending} deals with the bending of light in the strong deflection limit by the naked singularity, 
which is followed by a study of different observables in this scenario in Sec. \ref{sec:observables}. We then conclude 
with a summary and future directions in Sec. \ref{sec:summary}.
Throughout this work, we have used $c=G=1$ units, unless otherwise specified.


\section{Properties of the Naked Singularity Spacetime}
\label{sec:properties}

We start with a static, spherically symmetric spacetime, recently proposed by Joshi {\it et} al. in Ref. \cite{Joshi-2020}, 
with the following line element expressed in $(t,r,\theta,\phi)$ coordinates as
\begin{equation}
ds^2=-\frac{dt^2}{\left(1+\frac{M}{r}\right)^2}+\left(1+\frac{M}{r}\right)^2dr^2+r^2d\theta^2+r^2\sin^2\theta d\phi^2,
\label{eq:NS-metric}
\end{equation}
where $M$ is the ADM mass of the spacetime. The metric components $g_{tt}$ and $g_{rr}$ can be expanded in powers of $\frac{M}{r}$ as
\begin{align}
\nonumber g_{tt}&=-\left(1+\frac{M}{r}\right)^{-2}=-\left[1-\frac{2M}{r}+3\left(\frac{M}{r}\right)^2-~\cdots\right], \\ g_{rr}&=\left(1+\frac{M}{r}\right)^{2}=\frac{1}{\left(1+\frac{M}{r}\right)^{-2}}=\frac{1}{\left[1-\frac{2M}{r}+3\left(\frac{M}{r}\right)^2-~\cdots\right]}~.
\end{align}
Therefore, in the large-$r$ limit, it resembles the Schwarzschild spacetime, and in the asymptotically infinite $(r\to\infty)$ limit, 
it reduces to the flat Minkowski spacetime. Expressions of the Ricci and Kretschmann scalars of this spacetime appear in Ref. \cite{Joshi-2020}, 
and it can be shown that both diverge at $r=0$, which corresponds to a curvature singularity of this spacetime at $r=0$. 
Moreover, the $g_{rr}$ term in Eq. (\ref{eq:NS-metric}) is always finite and positive for the whole range of $r>0$. 
Therefore, the spacetime does not contain any event or absolute horizons, and hence the spacetime represented by 
the metric in Eq. (\ref{eq:NS-metric}) contains a globally naked singularity at $r=0$. It is shown in Ref. \cite{Joshi-2020} that the 
spacetime is seeded by an anisotropic fluid, which satisfies all the required energy conditions. It is also shown that the 
equation-of-state parameter of the anisotropic fluid tends to the value $-1/3$ as $r\to0$, and it becomes $1/3$ in the limit as $r$ tends to infinity.

\begin{figure}[h]
\centering
\includegraphics[scale=0.5]{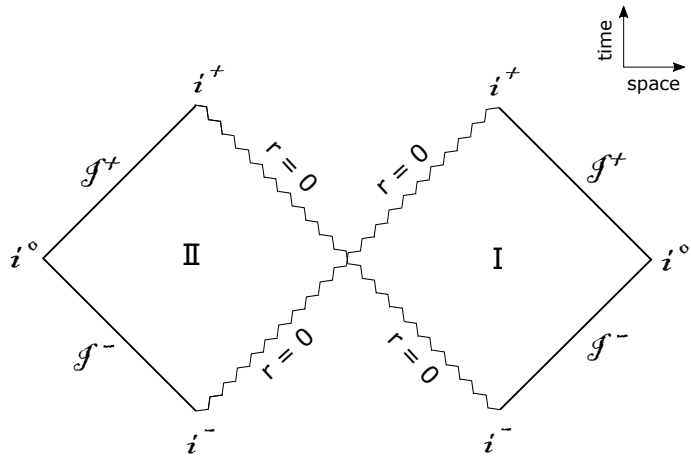}
\caption{Carter-Penrose diagram of the spacetime represented by the metric in Eq. (\ref{eq:NS-metric}). There are two different regions in the diagram, denoted by I and II. They are completely identical, but causally disconnected. The corresponding past null infinity $(\mathscr{I}^{-})$, future null infinity $(\mathscr{I}^{+})$, past timelike infinity $(i^{-})$, future timelike infinity $(i^+)$, and spacelike infinity $(i^0)$ are shown. The surface $r=0$ which represents the singularity is shown by the zigzagged line.}
\label{fig:CP-diagram}
\end{figure}

It is important to study the causal structure of the spacetime. These are best represented by the corresponding maximally extended spacetime diagrams, or the Carter-Penrose (CP) diagrams. Figure \ref{fig:CP-diagram} shows the CP diagram of the spacetime 
of Eq. (\ref{eq:NS-metric}).
The diagram consists of two distinct regions, denoted by I and II. They are identical copies of each other, but causally disconnected. The different kinds of infinities are denoted by the following standard notations: $\mathscr{I}^{\pm}$ denote the future and past null infinities, respectively; $i^{\pm}$ denote the future and past timelike infinities, respectively; and $i^0$ denotes the spacelike infinity. The hypersurface $r=0$ representing the singularity is denoted by the zigzagged line, corresponding to lightlike geodesics. As a result, the singularity at $r=0$ will be lightlike or null in nature. Therefore, in addition to being globally naked, the singularity is also lightlike, which is an interesting and novel feature of this spacetime.


\section{Characteristics of null geodesics}
\label{sec:photon-motion}
Let us now consider the nature of null geodesics in this spacetime. Since the spacetime is spherically symmetric, we can choose $\theta=\pi/2$ without any loss of generality. Moreover, it being a static, spherically symmetric spacetime, it admits two constants of motion--- namely, the energy $E$ and the $z$ component of angular momentum $L$. The corresponding $t$ and $\phi$ geodesic equations for photon motion are given by
\begin{equation}
\dot{t}=\left(1+\frac{M}{r}\right)^2E~, \quad \text{and} \quad \dot{\phi}=\frac{L}{r^2}~,
\label{eq:t-phidot}
\end{equation}
where the overdot represents a derivative with respect to the affine parameter along a null geodesic. From the normalization of the 
four-velocity of photons $(u^{\mu}u_{\mu}=0)$, we have
\begin{equation}
\dot{r}^2+\frac{L^2}{r^2\left(1+\frac{M}{r}\right)^2}=E^2 \quad \text{or,} \quad \dot{r}^2+V_{\text{eff}}(r)=E^2,
\label{eq:rdot}
\end{equation}
where $V_{\text{eff}}(r)=\frac{L^2}{\left(r+M\right)^2}$ is the effective potential for photon motion. At the turning point $(r_{\text{tp}})$, we have $\dot{r}=0$, which yields from Eq. (\ref{eq:rdot})
\begin{equation}
V_{\text{eff}}(r_{\text{tp}})=\frac{L^2}{\left(r_{\text{tp}}+M\right)^2}=E^2 \quad \text{or,} \quad \frac{L}{E}=b(r_{\text{tp}})=\sqrt{\left( r_{\text{tp}}+M\right)^2}=\left(r_{\text{tp}}+M\right),
\label{eq:impact-param}
\end{equation}
where $b=L/E$ represents the impact parameter of a light ray, which is a constant of motion for that ray.
\begin{figure}[h]
\centering
\subfigure[~Effective Potential]{\includegraphics[scale=1]{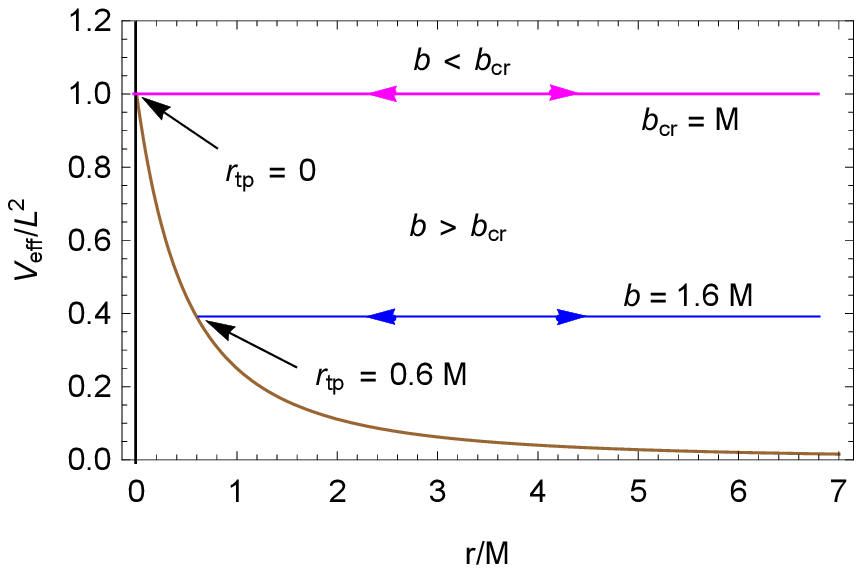}\label{fig:EffPot}}\hspace{0.1cm}
\subfigure[~Trajectories and shadow boundaries]{\includegraphics[scale=0.75]{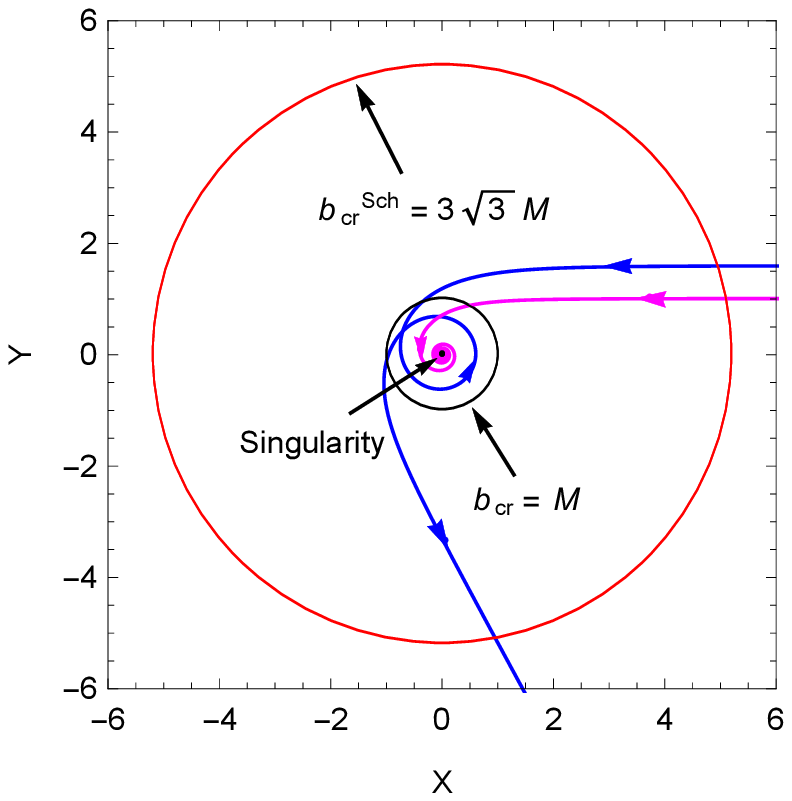}\label{fig:trajectory}}
\caption{(a) Plots of effective potential $V_{\text{eff}}$ (in units of $L^2$) as functions of $r$ (in units of $M$). (b) The associated trajectories of light. In these plots, we have shown two sets of rays. The blue ray has an impact parameter $b>b_{\text{cr}}$ ($b=1.6M$ to be precise), and the magenta ray corresponds to a ray having an impact parameter $b<b_{\text{cr}}=M$. The boundary of the shadow region is shown by the black circle in (b). We have also shown the shadow boundary of the Schwarzschild spacetime with a red circle in (b) to illustrate the relative sizes of the shadows in these two spacetimes. These plots are in agreement with the corresponding plots (Figs. 1(e) and 1(f)) obtained in Ref. \cite{Joshi-2020}.}
\label{fig:pot-trajectory}
\end{figure}
In Fig. \ref{fig:pot-trajectory}, we have plotted the effective potential and the corresponding light trajectories of  
the spacetime of Eq. (\ref{eq:NS-metric}). It is interesting to note from Fig. \ref{fig:EffPot} that $V_{\text{eff}}(r)$ is finite everywhere and does not contain any extremum within the whole range of $r\in\left.\left[0,\infty\right.\right)$. The value of $V_{\text{eff}}$ at the location of the singularity --- i.e., at $r=0$ --- is given by $\left.V_{\text{eff}}\right\rvert_{r=0}=\frac{L^2}{M^2}$. Remember that the maximum of $V_{\text{eff}}(r)$ corresponds to the position of what is known as a photon sphere (in spherically symmetric spacetimes), which represents a spherical surface consisting of unstable light rings. The most significant feature of such a photon sphere is that light coming from a distant source either crosses the surface of the photon sphere and spirals into it without ever coming out of the surface, or goes very close to it, takes a number of turns around it, and finally moves out towards infinity. As a result, the sky of a distant observer gets divided into two distinct regions: a completely dark region devoid of light (known as the shadow region), and a bright region surrounding the shadow. Therefore, the photon sphere, in this case, plays the most significant role in producing the shadow region.

Since $V_{\text{eff}}(r)$ of the naked singularity spacetime does not have any extremum, as can be seen from Fig. \ref{fig:EffPot}, it does not possess any photon sphere, and naively we do not expect it to produce a shadow. Interestingly however, as shown in Ref. \cite{Joshi-2020}, this spacetime indeed produces a shadow. From Eq. (\ref{eq:impact-param}), the value of the impact parameter $b$, for which a light ray has a turning point just at the singularity, can be obtained as $b\rvert_{r_{\text{tp}}=0}=M$. This corresponds to the critical ray forming the boundary of the shadow. In Fig. \ref{fig:trajectory}, we have shown two sets of light rays --- one representing a ray having impact parameter $b=1.6M>b_{\text{cr}}$ (blue ray), and the other one having impact parameter $b<b_{\text{cr}}=M$ (magenta ray). We can see that any ray of light having an impact parameter $b>M$ will take a turn at some finite $r_{\text{tp}}>0$. On the contrary, for light rays with impact parameters $b<M$, their geodesics will be incomplete and will be terminated at the singularity. Therefore, all such rays coming from a distant source are captured by the singularity, producing the shadow. It is worth noting here that, in the case of the Schwarzschild spacetime having the same ADM mass $M$, the corresponding critical impact parameter producing a shadow is given by $b_{\text{cr}}^{\text{Sch}}=3\sqrt{3}M$. Hence, the size of the shadow of the naked singularity spacetime of Eq. (\ref{eq:NS-metric}) is smaller than that of the Schwarzschild shadow. This feature has also been reported in Ref. \cite{Joshi-2020} in the case of a spherical accretion by the Schwarzschild black hole and the naked singularity under consideration. We also have shown the boundaries of the shadow regions for both naked singularity (black circle) and Schwarzschild (red circle) spacetimes in Fig. \ref{fig:trajectory} for illustrative purposes.


\section{Bending of light in the strong field limit}
\label{sec:strong-bending}
An important property of a spacetime having a photon sphere is that as light rays come very close to the photon sphere, they take a number of turns before reaching the observer. As a result, the bending angle of light becomes large and diverges at the location of the photon sphere. Since, in the present naked singularity spacetime, the shadow is produced without a photon sphere, it will be interesting to study how the light bending angle behaves in such a scenario. For this purpose, let us rearrange Eq. (\ref{eq:rdot}) to the form
\begin{equation}
\dot{r}=\pm\sqrt{E^2-\frac{L^2}{r^2\left(1+\frac{M}{r}\right)^2}}=\pm\sqrt{E^2-\frac{L^2}{\left(r+M\right)^2}}.
\label{eq:rdot-new}
\end{equation}
Combining the above equation and the second expression of Eq. (\ref{eq:t-phidot}), we obtain
\begin{equation}
\frac{d\phi}{dr}=\frac{\dot{\phi}}{\dot{r}}=\pm\frac{L}{r^2\sqrt{E^2-\frac{L^2}{\left(r+M\right)^2}}}=\pm\frac{b}{r^2\sqrt{1-\frac{b^2}{\left(r+M\right)^2}}}.
\label{eq:dphidr}
\end{equation}
Again, from Eq. (\ref{eq:impact-param}), we know $b(r_{\text{tp}})=\left(r_{\text{tp}}+M\right)$. Therefore, Eq. (\ref{eq:dphidr}) can be rewritten as
\begin{equation}
\frac{d\phi}{dr}=\pm\frac{\left(r_{\text{tp}}+M\right)}{r^2\sqrt{1-\frac{\left(r_{\text{tp}}+M\right)^2}{\left(r+M\right)^2}}}=\pm\frac{\left(r_{\text{tp}}+M\right)(r+M)}{r^2\sqrt{r^2+2\left(r-r_{\text{tp}}\right)M-r_{\text{tp}}^2}},
\end{equation}
whose integration gives
\begin{equation}
\phi(r)=\phi_{\infty}\pm\int_{r}^{\infty}\frac{\left(r_{\text{tp}}+M\right)(r+M)}{r^2\sqrt{r^2+2\left(r-r_{\text{tp}}\right)M-r_{\text{tp}}^2}}~dr
\end{equation}
where $\phi_{\infty}$ is the initial azimuthal angle made by the light ray when it originates from the source asymptotically. 
Therefore, the true angle of deflection of light due to the gravitational field produced by the naked singularity is given by
\begin{equation}
\alpha(r_{\text{tp}})=2~|\phi(r_{\text{tp}})-\phi_{\infty}|-\pi~~,~~{\rm i.e.,}~~
\alpha(r_{\text{tp}})=2\int_{r_{\text{tp}}}^{\infty}\frac{\left(r_{\text{tp}}+M\right)(r+M)}{r^2\sqrt{r^2+2\left(r-r_{\text{tp}}\right)M-r_{\text{tp}}^2}}~dr-\pi.
\label{eq:bending-int}
\end{equation}
The integration in Eq. (\ref{eq:bending-int}) is elementary, and we obtain the analytic expression of $\alpha(r_{\text{tp}})$ as
\begin{equation}
\alpha(r_{\text{tp}})=\frac{2M\left(r_{\text{tp}}+M\right)}{r_{\text{tp}}\left(r_{\text{tp}}+2M\right)}+\frac{4\left(r_{\text{tp}}+M\right)^3}{r_{\text{tp}}^{3/2}\left(r_{\text{tp}}+2M\right)^{3/2}}\arctan\left(\sqrt{1+\frac{2M}{r_{\text{tp}}}}~\right)-\pi.
\label{eq:bending-rtp}
\end{equation}
Equivalently, the bending angle can also be obtained in terms of the impact parameter $b$, and it is given by
\begin{equation}
\alpha(b)=\frac{2Mb}{\left(b^2-M^2\right)}+\frac{4b^3}{\left(b^2-M^2\right)^{3/2}}\arctan\left(\sqrt{\frac{b+M}{b-M}}~\right)-\pi.
\label{eq:bending-b}
\end{equation}
As we can see from Eqs. (\ref{eq:bending-rtp}) and (\ref{eq:bending-b}), the bending angle diverges in the limit of the critical impact parameter --- i.e., $b\to b_{\text{cr}}=M$ or equivalently, in the limit $r_{\text{tp}}\to0$. To find out the nature of the divergence of $\alpha$ in these limiting cases, we first need to use the expansion of $\arctan(x)$ for the $x\ge1$ case, given as
$\arctan(x)=\frac{\pi}{2}-\frac{1}{x}+\frac{1}{3x^3}-\frac{1}{5x^5}+\cdots~$, for $x\ge1$.
Therefore, the expressions of $\alpha(r_{\text{tp}})$ and $\alpha(b)$ can be expanded, respectively, as
\begin{eqnarray}
\alpha(r_{\text{tp}})&=&\left[\frac{2\pi(r_{\text{tp}}+M)^3}{r_{\text{tp}}^{3/2}(r_{\text{tp}}+2M)^{3/2}}-\frac{2(r_{\text{tp}}+M)(2r_{\text{tp}}+3M)}{(r_{\text{tp}}+2M)^2}+\frac{4(r_{\text{tp}}+M)^3}{3(r_{\text{tp}}+2M)^3}-\frac{4r_{\text{tp}}(r_{\text{tp}}+M)^3}{5(r_{\text{tp}}+2M)^4}+\cdots\right]-\pi~,
\nonumber\\
\alpha(b)&=&\left[\frac{2\pi(b/M)^3}{\left(\frac{b^2}{M^2}-1\right)^{3/2}}-\frac{2(b/M)\left(\frac{2b}{M}+1\right)}{\left(\frac{b}{M}+1\right)^2}+\frac{4(b/M)^3}{3\left(\frac{b}{M}+1\right)^3}-\frac{4(b/M)^3\left(\frac{b^2}{M^2}-1\right)}{5\left(\frac{b}{M}+1\right)^5}+\cdots\right]-\pi~.
\end{eqnarray}
Hence, the divergences of $\alpha(r_{\text{tp}})$ in the limit $r_{\text{tp}}\to0$, and $\alpha(b)$ in the limit $b\to b_{\text{cr}}=M$, respectively, take the form
\begin{eqnarray}
\lim_{r_{\text{tp}}\to0}\alpha(r_{\text{tp}})&=&\frac{\pi}{\sqrt{2}}\left(\frac{M}{r_{\text{tp}}}\right)^{3/2}-\frac{4}{3}-\pi+\mathcal{O}[r_{\text{tp}}]~,
\label{eq:bending-rtp-limit}
\\
\lim_{b\to b_{\text{cr}}}\alpha(b)&=&\frac{(\pi/\sqrt{2})}{\left(\frac{b}{b_{\text{cr}}}-1\right)^{3/2}}-\frac{4}{3}-\pi+\mathcal{O}\left[\left(\frac{b}{b_{\text{cr}}}-1\right)\right]~.
\label{eq:bending-b-limit}
\end{eqnarray}
From Eqs. (\ref{eq:bending-rtp-limit}) and (\ref{eq:bending-b-limit}), we observe that the nature of the divergence of $\alpha$ in the limits $b\to b_{\text{cr}}=M$ or $r_{\text{tp}}\to0$ is nonlogarithmic and is of polynomial type, which is a characteristic feature of this naked singularity. 
For completeness, let us recapitulate the different kinds of divergences of the bending angle in different situations.
\begin{itemize}
\item[] {\bf Case 1:} If a spacetime contains only a photon sphere $(r_{\text{ph}})$, which corresponds to the critical impact parameter $b=b_{\text{ph}}$, the bending angle diverges in the limit $r\to r_{\text{ph}}$ or $b\to b_{\text{ph}}$ as the light ray approaches the photon sphere from the $b>b_{\text{ph}}$ side. The corresponding divergence is logarithmic which was first shown by Bozza in Ref. \cite{Bozza-2002}, and later refined by Tsukamoto in Ref. \cite{Tsukamoto-2017}. If we write the metric of a general static, spherically symmetric spacetime as
\begin{equation}
ds^2=-A(r)dt^2+B(r)dr^2+C(r)(d\theta^2+\sin^2\theta d\phi^2),
\label{eq:metric-gen}
\end{equation}
then the corresponding formula for the bending angle in the limit $b\to b_{\text{ph}}$ will be given by \cite{Tsukamoto-2017}
\begin{equation}
\alpha(b)=-\bar{a}\log\left(\frac{b}{b_{\text{ph}}}-1\right)+\bar{b}+\mathcal{O}[(b-b_{\text{ph}})\log(b-b_{\text{ph}})],
\end{equation}
where the expressions of $\bar{a}$ and $\bar{b}$ are
\begin{equation}
\bar{a}=\left.\sqrt{\frac{2AB}{C''A-CA''}}~\right\rvert_{r_{\text{ph}}}, \quad \bar{b}=\left.\bar{a}\log\left[r^2\left(\frac{C''}{C}-\frac{A''}{A}\right)\right]\right\rvert_{r_{\text{ph}}}+I_R(r_{\text{ph}})-\pi,
\end{equation}
with a prime denoting a derivative with respect to $r$, and $I_R(r_{\text{ph}})$ is a constant that can be found in Ref. \cite{Tsukamoto-2017}.

\item[] {\bf Case 2:} If a spacetime contains both a photon $(r_{\text{ph}})$ and an antiphoton $(r_{\text{aph}})$ sphere,\footnote{By an antiphoton sphere, we refer to the minimum of the effective potential, just like a photon sphere corresponds to the maximum of the effective potential.}  the bending angle diverges in the limit $r\to r_{\text{ph}}$ or $b\to b_{\text{ph}}$ from both the $b>b_{\text{ph}}$ and the $b<b_{\text{ph}}$ sides. The nature of divergence due to the photon sphere --- i.e., from the $b>b_{\text{ph}}$ side --- is the same as described in Case 1. On the contrary, though the divergence due to the antiphoton sphere --- i.e., from the $b<b_{\text{ph}}$ side --- is also logarithmic in nature, its functional dependence is different from Case 1. The scenario has been discussed in detail in Ref. \cite{Shaikh-2019c}, and the corresponding formula of divergence from the $b<b_{\text{ph}}$ side is obtained as \cite{Shaikh-2019c}
\begin{equation}
\alpha(b)=-\bar{a}\log\left(\frac{b_{\text{ph}}^2}{b^2}-1\right)+\bar{b}+\mathcal{O}[(b_{\text{ph}}^2-b^2)\log(b_{\text{ph}}^2-b^2)],
\end{equation}
where the expressions of $\bar{a}$ and $\bar{b}$, in this case, are
\begin{equation}
\bar{a}=\left.2\sqrt{\frac{2AB}{C''A-CA''}}~\right\rvert_{r_{\text{ph}}}, \quad \bar{b}=\left.\bar{a}\log\left[2r^2\left(\frac{C''}{C}-\frac{A''}{A}\right)\left(\frac{r}{r_c}-1\right)\right]\right\rvert_{r_{\text{ph}}}+I_R(r_c)-\pi,
\end{equation}
and $r_c~(<r_{\text{ph}})$ is the radius for which $V_{\text{eff}}(r_c)=V_{\text{eff}}(r_{\text{ph}})$ \cite{Shaikh-2019c}. Therefore, it is evident that the logarithmic divergences due to a photon sphere (Case 1) and an antiphoton sphere (Case 2) are quite different, which results in some important observational consequences as described in detail in Ref. \cite{Shaikh-2019c}.

\item[] {\bf Case 3:} There exists a critical situation in between the previous two cases--- namely, when the photon and antiphoton spheres merge together to form a point of inflection in the effective potential. Let us denote this point of inflection as $r_0~(=r_{\text{ph}}=r_{\text{aph}})$, and the corresponding impact parameter as $b_0~(=b_{\text{ph}})$. This scenario has been recently addressed by Tsukamoto in Ref. \cite{Tsukamoto-2020} for the Damour-Solodukhin wormhole. The importance of this analysis is that the corresponding divergence of the bending angle turns out to be nonlogarithmic, given by the following expression \cite{Tsukamoto-2020}:
\begin{equation}
\alpha(b)=\frac{\bar{a}}{\left(\frac{b}{b_0}-1\right)^{1/4}}+\bar{b}+\mathcal{O}\left[\left(\frac{b}{b_0}-1\right)^{3/4}\right],
\end{equation}
where $\bar{a}$ and $\bar{b}$ are given by
\begin{equation}
\bar{a}=2^{7/4}~3^{-1/4}=2.5558, \quad \bar{b}=-\frac{4}{3}+I_R(r_0)-\pi=-2.1078.
\end{equation}
\end{itemize}
In the present analysis, we have seen from Eq. (\ref{eq:bending-rtp-limit}) or Eq. (\ref{eq:bending-b-limit}) that the spacetime 
of Eq. (\ref{eq:NS-metric}) also admits nonlogarithmic divergence, as in Case 3. Importantly, the nature of divergence is different from that of Case 3. In this case, the corresponding bending angle [from Eq. (\ref{eq:bending-b-limit})] turns out to be 
\begin{equation}
\alpha(b)=\frac{\bar{a}}{\left(\frac{b}{b_{\text{cr}}}-1\right)^{3/2}}+\bar{b}+\mathcal{O}\left[\left(\frac{b}{b_{\text{cr}}}-1\right)\right]~,
\end{equation}
where $\bar{a}=\frac{\pi}{\sqrt{2}}=2.2214$ and $\bar{b}=-\frac{4}{3}-\pi=-4.4749$, with $b_{\text{cr}}=M$.

For a spacetime having a photon sphere, a light ray with an impact parameter exactly equal to its critical value will arrive at the surface of the sphere tangentially, and in principle, remain there. Naturally, such a ray will have an infinite bending angle. For rays having impact parameters just above the critical value, their turning points will be close to the photon sphere, and they will make a large number of rotations before escaping to the observer. In other words, as the value of the impact parameter approaches its critical value from a larger one, the corresponding turning points of the rays will be closer to the photon sphere, and the closer the turning points are towards the photon sphere, the more the bending angle will be. Therefore, the divergence of the bending angle at the photon sphere occurs naturally. On the other hand, the notion of such a divergence of the bending angle without a photon sphere is interesting and deserves a careful analysis.
\begin{figure}[h]
\centering
\subfigure[]{\includegraphics[scale=0.8]{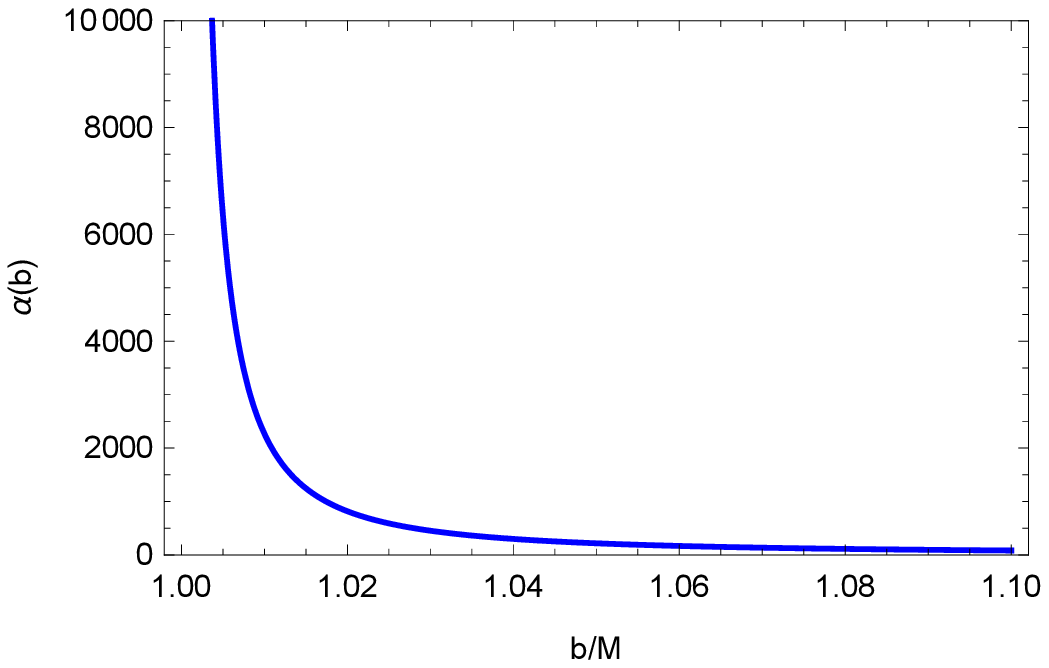}\label{fig:bending-b}}\hspace{0.2cm}
\subfigure[]{\includegraphics[scale=0.8]{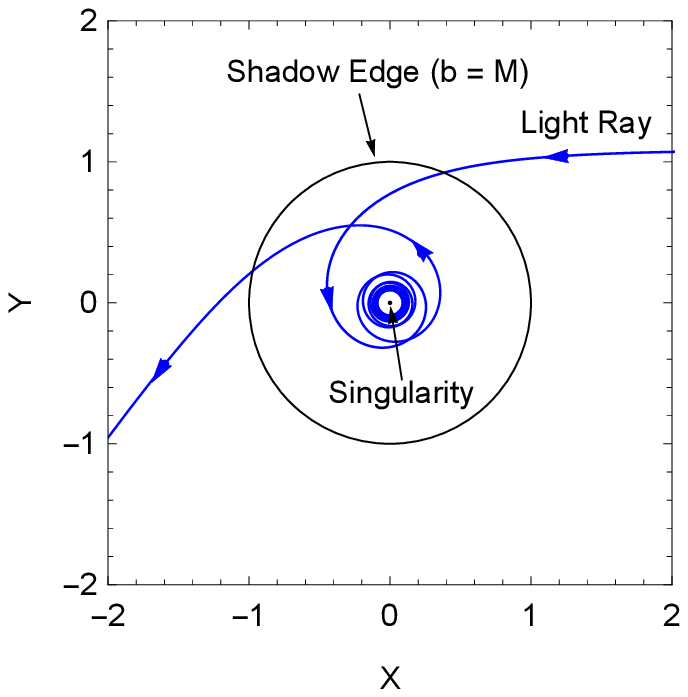}\label{fig:bending-trajectory}}
\caption{(a) Bending angle $\alpha$ as a function of the impact parameter $b$ (in units of $M$). (b) A representative trajectory of light having impact parameter just greater than its critical value.}
\label{fig:bending-angle}
\end{figure}

In Fig. \ref{fig:bending-b}, we have plotted the bending angle $(\alpha)$ as a function of the impact parameter $(b)$, and in Fig. \ref{fig:bending-trajectory}, the trajectory of a typical ray of light having an impact parameter just greater than the critical value for the present spacetime has been shown. We can see from Fig. \ref{fig:bending-b} that $\alpha(b)$ diverges as $b\to b_{\text{cr}}=M$. Figure \ref{fig:bending-trajectory} is particularly important for us to understand the reason of this divergence. Since the singularity is lightlike, light rays originating from a distant source can come very close to it--- or  in fact, in the critical case, they can even skim along the singularity without being truly captured by it. So, even the critical ray, after reaching the singularity and taking a turn there, can in principle reach the observer. In doing so, such rays take many turns (infinite in the critical case) around the singularity before reaching the observer to produce a large deflection. Only the rays having impact parameters less than the critical value are captured by the singularity, and the corresponding geodesics terminate without reaching the observer. Therefore, the nature of this naked singularity itself causes the bending angle to diverge.


\section{Observables in Strong lensing}
\label{sec:observables}

\begin{figure}[h]
\centering
\includegraphics[scale=2.0]{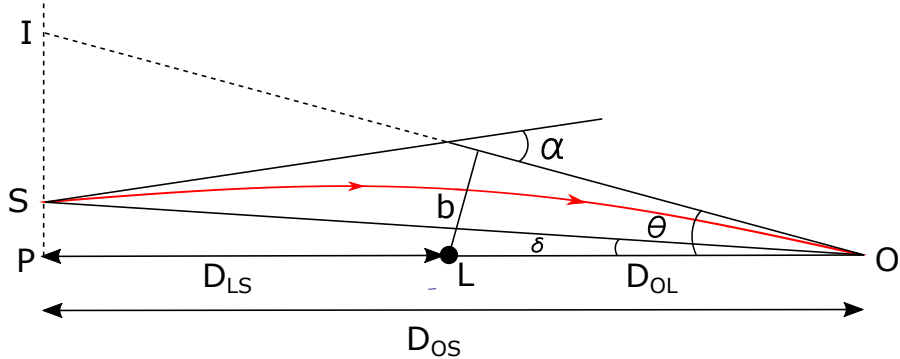}
\caption{Lens diagram. Light is emitted by the source (S), passes through the gravitational field of the lensing object (L), and then is received by the observer (O). Both the source and the observer are situated at the asymptotically flat region. OLP is the optic axis, $D_{\text{OS}}$, $D_{\text{LS}}$, and $D_{\text{OL}}$ represent the observer-source, lens-source, and observer-lens distances, respectively. The angular positions of the source (S) and the image (I) with respect to the optic axis are denoted by the angles $\delta$ and $\theta$, respectively. The impact parameter and the bending angle are represented by $b$ and $\alpha$, respectively.}
\label{fig:lens-diagram}
\end{figure}

In this section, we consider various observables of gravitational lensing, in the strong deflection limit. The lens diagram is 
shown in Fig. \ref{fig:lens-diagram}.
A ray of light is emitted by the source S and received by the observer O, encountering the gravitational field produced by the lensing object L and then producing the image I. Both the source and the observer are situated at an asymptotically flat region, so that the bending is due to 
the lens alone. The corresponding lens equation can be expressed as \cite{Virbhadra-2000}
\begin{equation}
\tan\delta=\tan\theta-\frac{D_{\text{LS}}}{D_{\text{OS}}}\left[\tan\theta+\tan(\alpha-\theta)\right],
\label{eq:lens-exact}
\end{equation}
where $\delta$ and $\theta$ are the angular positions of the source and the image, respectively, with respect to the optic axis, as shown in 
Fig. \ref{fig:lens-diagram}. $D_{\text{OS}}$ and $D_{\text{LS}}$ represent the observer-source distance and the lens-source distance, respectively, and $\alpha$ is the total bending angle. The impact parameter $b$ can be written in terms of $\theta$ as
\begin{equation}
b=D_{\text{OL}}\sin\theta=\kappa\sin\theta \quad\quad\quad\quad [\text{where}~\kappa=D_{\text{OL}}],
\label{eq:b-exact}
\end{equation}
with $D_{\text{OL}}$ being the observer-lens distance. The radial $(\mu_r)$, tangential $(\mu_t)$, and total $(\mu)$ magnifications of the image are expressed as
\begin{equation}
\mu_r=\left(\frac{\partial\delta}{\partial\theta}\right)^{-1}, \quad \mu_t=\left(\frac{\sin\delta}{\sin\theta}\right)^{-1}, \quad \text{and} \quad \mu=\mu_t\cdot\mu_r=\left(\frac{\sin\delta}{\sin\theta}~\frac{\partial\delta}{\partial\theta}\right)^{-1}.
\label{eq:magnification-exact}
\end{equation}
We shall use the three relations of Eq. (\ref{eq:magnification-exact}) for the rest of our calculations. 
Since we have an analytic formula for the bending angle, we can find out the expressions of the quantities in analytic forms. Using Eqs. (\ref{eq:bending-b}) and (\ref{eq:b-exact}), we get
\begin{equation}
\alpha(\theta)=\frac{2M\kappa\sin\theta}{\left(\kappa^2\sin^2\theta-M^2\right)}+\frac{4\kappa^3\sin^3\theta}{\left(\kappa^2\sin^2\theta-M^2\right)^{3/2}}\arctan\left(\sqrt{\frac{\kappa\sin\theta+M}{\kappa\sin\theta-M}}~\right)-\pi.
\end{equation}
Using this $\alpha(\theta)$ in Eq. (\ref{eq:lens-exact}), we obtain the source position $(\delta)$ as a function of the image position $(\theta)$, and from $\delta(\theta)$, it is straightforward to evaluate the expressions of the magnifications as functions of $\theta$. Note from Eq. (\ref{eq:magnification-exact}) that, when $\delta=0$ --- i.e., when the source, lens and observer are perfectly aligned along the optic axis --- $\mu_t$ diverges. In this scenario, the images form perfect concentric rings in the observer's sky, which are known as Einstein rings (when the bending of light is less than $2\pi$), or relativistic Einstein rings (when light bends by more than $2\pi$). Moreover, the singularities in magnification in the image plane are known as critical curves, and the corresponding locations in the source plane are called caustics, which produce the critical curves. The singularities in $\mu_r$ correspond to radial critical curves (RCCs) in the image plane and radial caustics (RCs) in the source plane, and similarly, the singularities in $\mu_t$ represent tangential critical curves (TCCs) in the image plane and tangential caustics (TCs) in the source plane. Clearly, therefore, the rings represent the TCCs, and correspondingly, $\delta=0$ gives the TC.

Now, to obtain the angular positions of the rings, we set $\delta=0$ in Eq. (\ref{eq:lens-exact}) and find
\begin{equation}
\tan[\alpha(\theta)-\theta]=\frac{D_{\text{OS}}}{D_{\text{LS}}}\left(1-\frac{D_{\text{LS}}}{D_{\text{OS}}}\right)\tan\theta.
\end{equation}
In our calculation, we have considered $D_{\text{OL}}=D_{\text{LS}}=\frac{1}{2}D_{\text{OS}}$. Therefore, the above equation yields
\begin{equation}
\tan[\alpha(\theta)-\theta]=\tan\theta, \quad \implies \quad \alpha(\theta)=m\pi+2\theta \quad\quad [m=1,2,3, ...].
\end{equation}
Since only the rays which take one or more complete rotations around the lens during its travel from the source to the observer can reach the observer, only even values of $m$ will contribute to the ring formation in the observer's sky. Thus, the position of the $n$th ring $(\theta_n)$ can be conveniently determined from the points of intersections of the curves, $y_1(\theta_n)=\alpha(\theta_n)$ and $y_2(\theta_n)=2n\pi+2\theta_n$, with $n=1,2, ...$ . Even if we do not assume the condition $D_{\text{OL}}=D_{\text{LS}}=\frac{1}{2}D_{\text{OS}}$, we can still find out $\theta_n$ by similarly evaluating the points of intersections of the two curves, $y_1(\theta_n)=\tan[\alpha(\theta_n)-\theta_n]$ and $y_2(\theta_n)=\frac{D_{\text{OS}}}{D_{\text{LS}}}\left(1-\frac{D_{\text{LS}}}{D_{\text{OS}}}\right)\tan\theta_n$, with $n=1,2, ...$ . Note that $n=1$ represents the outermost ring, and as we increase $n$, the angular positions of the relativistic rings decrease, and the corresponding bending angle increases. The $n=\infty$ case represents the innermost critical ring which forms the edge of the shadow region, and its angular position is denoted by $\theta_{\infty}$. This ring is produced by the critical ray having the impact parameter $b=b_{\text{{cr}}}=M$.
\begin{figure}[h]
\centering
\subfigure[Radial Magnification]{\includegraphics[scale=0.9]{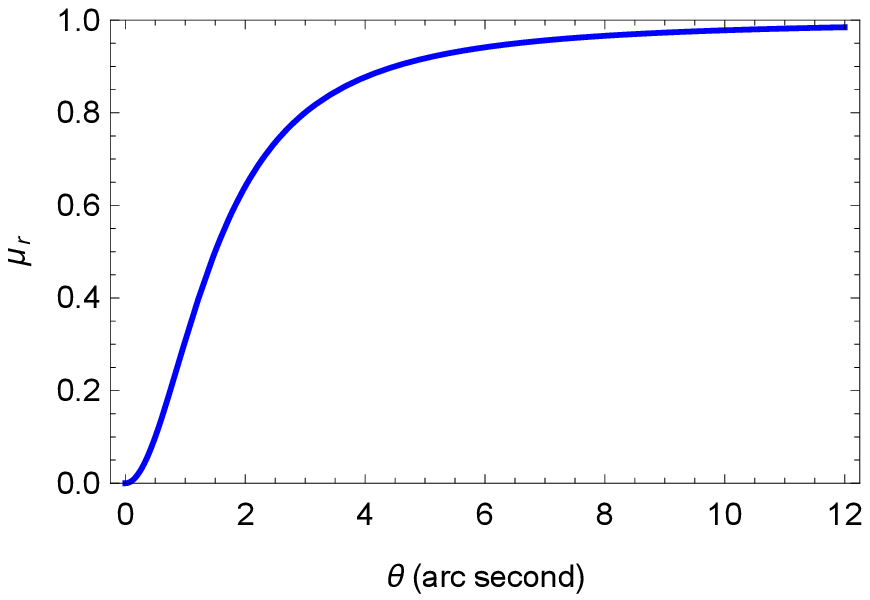}\label{fig:rad-mag}}\hspace{0.2cm}
\subfigure[Tangential Magnification]{\includegraphics[scale=0.87]{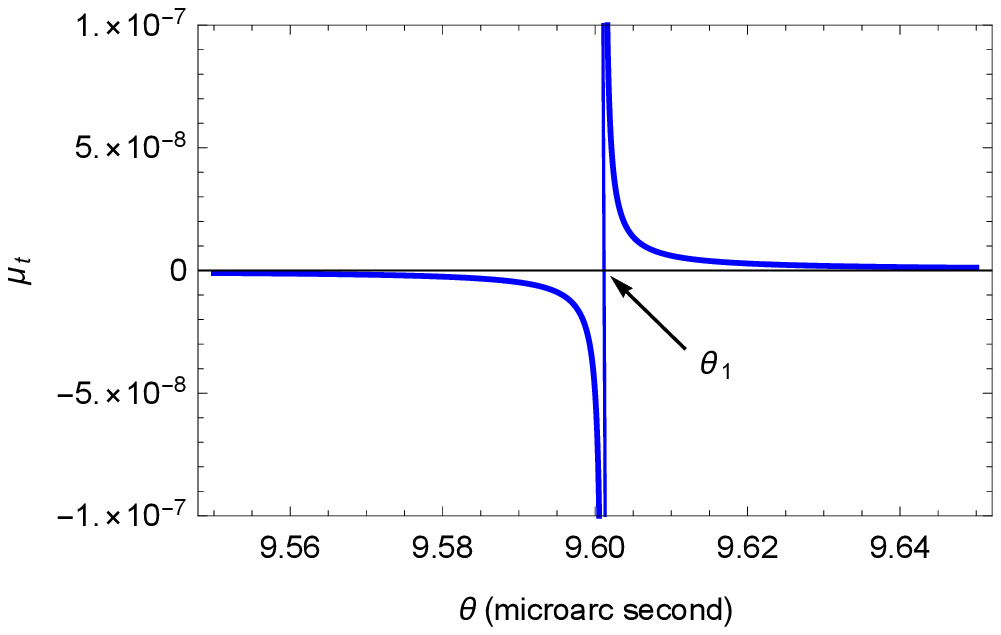}\label{fig:tan-mag}}
\caption{Radial $(\mu_r)$ and tangential $(\mu_t)$ magnifications as functions of the image angular position $(\theta)$. Here, (a) represents the variation of $\mu_r$, and (b) corresponds to the plot of $\mu_t$. In this case, the point of divergence of $\mu_t$ denotes the angular position of the first (outermost) relativistic Einstein ring. Here, $\theta$ is expressed in arcseconds in the left figure (radial magnification), and in micro-arcseconds in the right figure (tangential magnification).}
\label{fig:magnifications}
\end{figure}
In Fig. \ref{fig:magnifications}, we have plotted $\mu_r$ and $\mu_t$ of the images as functions of the image positions $(\theta)$. In the plots, the following parameter values are used : 
\[M=4.31\times10^6~ M_{\odot},~~ D_{\text{OL}}=D_{\text{LS}}=7.86~ \text{kpc}, \quad \text{and} \quad D_{\text{OL}}=\frac{1}{2}D_{\text{OS}}.\]
The above mass and distance correspond to the supermassive central object (considered as a black hole, Sgr A$^*$) of our Galaxy. Moreover, the unit of $\theta$ is expressed in arcseconds in the radial magnification plot, and in micro-arcseconds in the tangential magnification plot. We can see from Fig. \ref{fig:rad-mag} that the radial magnification never diverges, which signifies the absence of RCC or, correspondingly, RC in this spacetime. It can be shown that $\mu_r$ diverges only when $\frac{\partial\alpha(\theta)}{\partial\theta}>0$, whereas, in the present scenario, $\frac{\partial\alpha(\theta)}{\partial\theta}$ is always negative, as $\alpha(\theta)$ continuously increases with decreasing $\theta$. Therefore, the radial magnifications of the images always remain finite, ranging from $0$ to $1$. On the other hand, Fig. \ref{fig:tan-mag} shows that the tangential magnification $\mu_t$ (corresponding to $n=1$) diverges at the angular position $\theta_1$, which marks the position of the outermost relativistic Einstein ring.

To perform a comparative analysis between the Schwarzschild black hole and the naked singularity, the angular positions of a few relativistic Einstein rings (bending angle $>2\pi$) and the corresponding separations between two consecutive rings, defined as $s_n=\theta_n-\theta_{(n+1)}$, are tabulated in Table \ref{tab:image-parameters}.
\begin{table}[ht]
\centering
\caption{Angular positions of the Einstein ring $(\theta_{\text{E}})$ and a few relativistic Einstein rings $(\theta_n)$ (bending angle $>2\pi$) are shown for both the Schwarzschild black hole and the naked singularity spacetime. We have also shown the separation between two consecutive rings, $s_n=\theta_n-\theta_{n+1}$. Here, we have assumed the parameter values $M=4.31\times10^6~ M_{\odot}$, $D_{\text{OL}}=D_{\text{LS}}=7.86~ \text{kpc}$, and $D_{\text{OL}}=\frac{1}{2}D_{\text{OS}}$, which correspond to the supermassive central object (considered as a black hole, Sgr A$^*$) of our Milky Way Galaxy. The angles are expressed in micro-arcseconds. Note that $\theta_{\text{E}}$ has a value of the order of arcseconds.}
\label{tab:image-parameters}
\vspace{10pt}
\begin{spacing}{1.3}
\begin{tabular}{|c| c| c|}
\hline\hline
          					   &~~Schwarzschild~~    &~~Naked singularity~~ \\
          					   &					 &of metric in Eq. (\ref{eq:NS-metric}) \\
\hline
$\theta_{\text{E}}$            &$1.49745\times10^6$  &$1.49747\times10^6$ \\
\hline
$\theta_1$  				   &$28.280347$          &$9.60106$ \\
$\theta_2$   				   &$28.24494$           &$7.75320$ \\
$\theta_3$ 					   &$28.24484$		     &$7.10409$ \\
$\theta_4$  				   &$\cdots$             &$6.76498$ \\
$\theta_{\infty}$              &$28.24485$           &$5.43573$ \\
\hline
$s_1$       				   &$0.035407$           &$1.84786$ \\
$s_2$       				   &$0.00009$            &$0.64911$ \\
$s_3$        				   &$\cdots$             &$0.33911$ \\
$s_{50}$      				   &$\cdots$             &$0.00290$ \\
$~\theta_1-\theta_{\infty}~$   &$0.035497$           &$4.16533$ \\
\hline\hline
\end{tabular}
\end{spacing}
\end{table}
To this end, it should be noted that, apart from these relativistic rings, there exists one more ring well outside this relativistic ring system, for which the bending angle is very small. This additional ring is known as the Einstein ring denoted as $\theta_{\text{E}}$ in Table \ref{tab:image-parameters}. The angular positions (in arcseconds) of this ring for both the Schwarzschild black hole and the naked singularity are found to be almost equal. This signifies that, in the weak field limit, the naked singularity spacetime behaves like the Schwarzschild one, as stated earlier. 

\begin{figure}[H]
\centering
\includegraphics[scale=1.0]{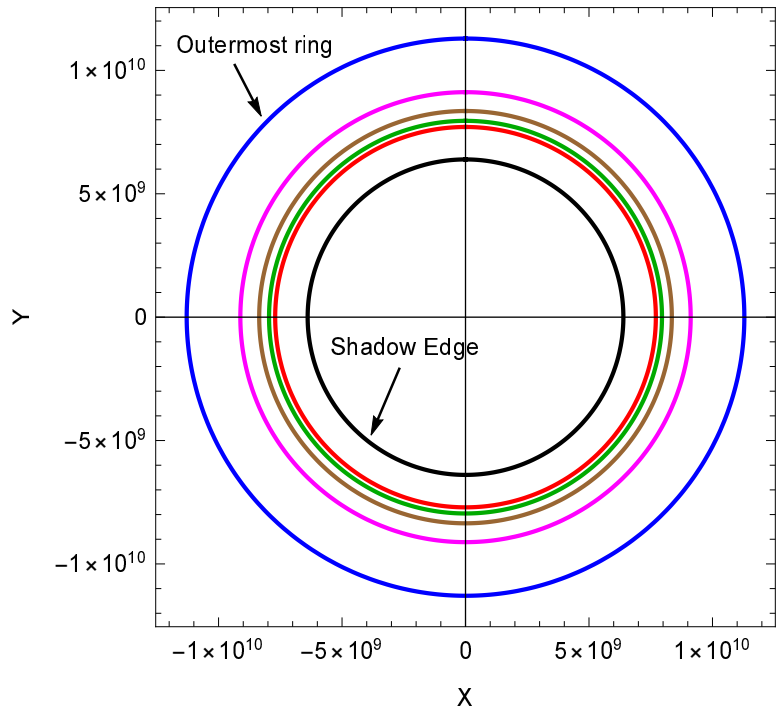}
\caption{Relativistic Einstein rings as seen in the observer's sky. The outermost ring is shown in blue, and the innermost one, forming the edge of the shadow, is shown in black. The second, third, fourth and fifth rings are denoted in magenta, brown, green, and red, respectively.}
\label{fig:rings}
\end{figure}

Things change dramatically in the strong deflection limit. As can be seen from Table \ref{tab:image-parameters}, the rings in Schwarzschild spacetime are very close to each other and converge quickly. In fact, the angular separation between the first and the last ring $(\theta_1-\theta_{\infty})$ is only $0.035497$ micro-arcseconds. From the values of angular positions of the rings in Schwarzschild spacetime, we see that only the first ring can be resolved from the rest. On the other hand, in the naked singularity spacetime, the rings are well separated from each other, and many rings can be individually resolved from the ring system before they finally become convergent. To visualize this feature, we have plotted, in Fig. \ref{fig:rings}, the first five relativistic rings, as well as the last one, which forms the edge of the shadow, as seen by the observer.
From the figure also, we observe that the rings are well separated, and therefore, it acts as the most crucial and prominent distinguishing feature of this naked singularity. These features may prove to be critical for the detection of such naked singularities in futuristic experiments.

\section{Summary}
\label{sec:summary}
Currently, observational signatures of black holes and horizonless objects attract much attention. 
The Event Horizon Telescope has opened up the possibility of experimentally probing gravity in its strongest regime, near the event horizon 
of a black hole or close to a singularity. These might provide important clues towards understanding the nature of quantum gravity. A number of studies have appeared in the recent literature about how to make observational distinctions between objects with and without event horizons. A broad consensus in  this regard is that sometimes the horizonless objects do possess distinguishing signatures from that of black holes, while in many cases, they are indistinguishable, and mimic black holes. In fact, it has recently been shown in Ref. \cite{Bambi-2019} that a Kerr superspinar can produce a shadow having a size and shape similar to the first image of the center of the M87$^*$ galaxy captured by the EHT. In this scenario, it is important and interesting to glean further insight into the differences of strong gravitational lensing between black holes and horizonless objects.  

In this paper, we have performed an extensive study of gravitational lensing in the strong deflection limit by a strongly naked null singularity proposed recently in Ref. \cite{Joshi-2020}. We have shown that the naked singularity is lightlike, and then we have discussed its lensing properties. We have
uncovered novel lensing phenomena that this spacetime might give rise to. First, we have shown that, in spite of being a strongly naked singularity, the bending angle of light diverges in a critical limit, which normally happens in a spacetime having a photon sphere. 
Moreover, we have found that the nature of this divergence is nonlogarithmic, contrary to black holes, and we have derived 
an analytic formula for the same. We have also analyzed various observables in this spacetime, and have found that the relativistic Einstein rings are well separated, and so, many of them can be easily resolved as compared to the Schwarzschild case. We believe that all these results are novel and add significantly to the existing literature. 

An immediate extension of this work would be to find a rotating version of such a singularity, and to study its characteristic features as far as strong gravitational lensing is concerned. We leave this for a future work.

\section{Acknowledgments}
The author thanks T. Sarkar and R. Shaikh for a preliminary reading of the manuscript, and for providing useful suggestions.

\end{document}